# The fragility of AI companionship: Ontological, structural, and normative uncertainty in human-AI relationships

Renwen Zhang[1]*, Lezi Xie[2]

[1] Wee Kim Wee School of Communication and Information, Nanyang Technological University, Singapore

[2] Department of Communication, University of California, Davis, USA

** Corresponding author: Renwen Zhang, renwen.zhang@ntu.edu.sg*

## Abstract

As generative AI chatbots become more personalized and emotionally responsive, they increasingly serve as companions, friends, and romantic partners. Yet these relationships are accompanied by significant uncertainty regarding AI's sentience, authenticity, and relational stability. Drawing on in-depth interviews with 25 users of AI companions, this study identifies three key forms of uncertainty in human-AI relationships: ontological uncertainty concerning the AI's nature and agency, structural uncertainty arising from platform control and system instability, and normative uncertainty regarding the legitimacy and boundaries of human-AI intimacy. Participants managed these uncertainties through information seeking, topic avoidance, expectation adjustment, and disengagement. This study extends interpersonal uncertainty theories to human-AI communication and contributes to HCI research by conceptualizing uncertainty as a socio-technical and relational phenomenon with socio-emotional implications. We discuss implications for designing safer AI companionship through contextual transparency, user control, update notice, and relational safeguards.

**Keywords:** Uncertainty; AI companionship; Human-AI interaction; AI ethics; Qualitative method

## Introduction

Recent advances in generative AI have transformed chatbots into social companions capable of simulating empathy and providing emotional support. People increasingly form bonds that resemble friendship or romance with AI chatbots, such as Replika and ChatGPT, blurring the line between human partner and machine (Brandtzaeg et al., 2022; Skjuve et al., 2021). Yet this blurred boundary gives rise to uncertainty about AI's ontological status (i.e., whether AI can think or feel), further compounded by the algorithmic opacity of these systems, which obscures how and why responses are generated (Liu, 2021; Pan et al., 2024). Such uncertainty can provoke frustration, erode trust, and hinder decision-making (Li and Zhang, 2024; Prabhudesai, et al., 2023). However, most studies about uncertainty in human-AI interaction focused on task-oriented contexts, such as decision-making support (Liu, 2021; Liu et al., 2023); less is known about the forms of uncertainty that emerge in socio-emotional interactions with AI, where users must navigate both the technical opacity and the emotional engagement with non-human agents (Banks, 2024; Zhang et al., 2025).

Interpersonal communication research offers a rich foundation for examining uncertainty in human-AI relationships. Uncertainty is defined as the degree of doubt or ambiguity individuals have in explaining and predicting their own or their partner's thoughts, feelings, and behaviors in social interactions (Berger and Calabrese, 1975; Berger and Bradac, 1982). Classic theories such as uncertainty reduction theory (Berger and Calabrese, 1975) and relational turbulence theory (Solomon et al., 2016) highlight its far-reaching relational, emotional, and communicative consequences. Uncertainty shapes the development of trust and intimacy, influences relationship maintenance, and affects emotional and mental well-being (Afifi and Reichert, 1996; Knobloch and Solomon, 2003; Theiss and Estlein, 2014). When faced with ambiguity, individuals are inclined to reduce uncertainty through a range of strategies, such as direct questioning, seeking information from third parties, and passive observation (Berger, 1995; Berger and Bradac, 1982).

However, these interpersonal theories are premised on assumptions of shared consciousness, intentionality, and emotional capacity, which do not hold in human-AI interactions. The algorithmic nature of AI agents introduces unique uncertainty about agency and sentience, largely absent from human-human relationships (Pan et al., 2024). Moreover, the sociotechnical infrastructure that governs AI companionship, such as system updates and content moderation policies, can unpredictably alter AI's persona or memory functions, disrupting emotional and relational continuity (Banks, 2024; De Freitas et al., 2024). These qualitatively distinct forms of uncertainty demand conceptual expansion of interpersonal theories to account for algorithmic mediation in relational life. To achieve this, it is imperative to adopt an inductive approach to understanding how uncertainty is lived and negotiated in everyday interactions with AI companions, which deepens our understanding of the dynamics of human-machine communication (HMC; Guzman and Lewis, 2020) and informs the design of AI systems that support users' relational satisfaction and emotional well-being.

This study investigates how uncertainty is perceived, experienced, and managed in human-AI relationships. Drawing on in-depth interviews with 25 users of AI companions, we examine the types of uncertainty that arise, their sociotechnical antecedents, and how users cope with or adapt to them. Our analysis identifies three distinct forms of uncertainty in human-AI relationships: ontological uncertainty (ambiguity about AI's agency and sentience), structural uncertainty (ambiguity about the sociotechnical infrastructure that sustains the relationship), and normative uncertainty (ambiguity about the social legitimacy and relational norms of human-AI intimacy). These uncertainties are shaped by factors such as algorithmic opacity, anthropomorphic design, platform instability, and the absence of social/relational norms. We also uncover four uncertainty management strategies, including information acquisition, topic avoidance, expectation adjustment, and disengagement.

This study makes three key contributions to communication and human-computer interaction (HCI) research. First, it identifies novel forms of uncertainty unique to human-AI interaction, extending classic interpersonal frameworks and enriching HMC scholarship. Second, it advances theoretical understanding of uncertainty as an ongoing communicative process rather than a static variable, revealing how socio-technical systems shape relational meaning-making. Third, it offers design implications for creating AI systems that are more transparent, trustworthy, and supporting users' well-being in the long run.

## Literature review

### Uncertainty of human-AI relationships

With advances in LLMs and generative AI, conversational agents (or chatbots) become increasingly adept at engaging with users socially and emotionally (Brandtzaeg et al., 2022; Skjuve et al., 2021). These natural, human-like conversations can foster the development of close relationships between humans and AI chatbots, spanning companionship, friendship, and romantic bonds (Croes and Antheunis, 2021; Li and Zhang, 2024; Skjuve et al., 2022), which resemble some key elements of interpersonal relationships like self-disclosure, social support, intimacy, and conflict (Li and Zhang, 2024; Liu and Sundar, 2018; Skjuve et al., 2021). However, human-AI relationships also raise considerable concerns. For example, emotional dependency on AI companions may lead to addiction, distress, and social withdrawal (Laestadius et al., 2024; Xie et al., 2023), and abrupt disruption of these attachments causes user frustration and emotional distress (Banks, 2024; Laestadius et al., 2024; Skjuve et al., 2022; Xie et al., 2023). Moreover, AI's lack of genuine reciprocity can weaken relationship quality and depth (Brandtzaeg et al., 2022; Croes and Antheunis, 2021; Skjuve et al., 2022). Privacy concerns also arise from extensive disclosure of personal sensitive information in human-AI interactions (Skjuve et al., 2022; Sullivan et al., 2023; Zhang et al., 2025). Algorithmic biases can generate inappropriate or misleading responses (Bender et al., 2021; Skjuve et al., 2022), potentially reinforcing gender and racial biases and unequal power structure (Depounti et al., 2023; Klonschinski and Kühler, 2021).

However, a crucial yet underexplored factor that may lead to socio-emotional harms is uncertainty. Uncertainty refers to individuals' lack of confidence in predicting, interpreting, or explaining their own and their partner's thoughts, feelings, behaviors, or relational dynamics (Berger and Calabrese, 1975). Studies also show that users of AI chatbots frequently encounter uncertainty regarding the AI's capability, privacy, and morality, which can erode trust, reduce user satisfaction, and undermine well-being (Banks, 2024; Croes and Antheunis, 2021; Pan et al., 2024; Zimmerman et al., 2023). Moreover, the anthropomorphic design of AI chatbots often make users question the AI's consciousness and intentionality (Pan et al., 2024), evoking feelings of unease, fear, or emotional discomfort (Li and Zhang, 2024). Unexpected software updates of AI companions can disrupt interaction patterns and relational dynamics, causing distress and emotional harm (Laestadius et al., 2024). For example, Replika's abrupt removal of erotic role-play feature in 2023 has stirred public outcry that eventually forced the company to restore this feature (Hanson and Bolthouse, 2024). Similarly, the abrupt shutdown of the AI companion Soulmate has induced enormous grief comparable to the loss of a loved one (Banks, 2024).

Despite its prevalence, uncertainty in human-AI interactions remains insufficiently theorized and empirically examined. Existing research on uncertainty in HCI has often focused on task-oriented human-AI interactions, examining ambiguity about the AI's accuracy, credibility, and influence on decision-making (Liu, 2021; Liu et al., 2023; Prabhudesai, et al., 2023). However, little is known about how uncertainty manifests in social and emotional interactions, where emotional connection—rather than accuracy or task efficiency—is the primary goal. One exception is Pan et al. (2024), who identified four types of uncertainty in human-AI relationships: technical, relational, ontological, and sexual uncertainty. While this typology offers valuable descriptive categories, it does not fully explain the underlying sources of uncertainty, nor does it explain how such uncertainty is managed in socio-emotional interactions with AI.

In parallel, research in Human-Computer Interaction, particularly within the Conversational User Interfaces (CUI) literature, has examined how users interpret and respond to ambiguity, breakdown, and inconsistency in conversational systems. For example, Lee et al. (2019) demonstrates how uncertainty and conversational breakdown shape users' perceptions of agency and relational expectations toward AI systems. Related studies also highlight how design features such as anthropomorphism, interaction style, and system transparency influence users' sensemaking, trust, and engagement (Khurana et al., 2021; Li et al., 2025; Liao and Vaughan, 2023). However, these insights have not been systematically integrated into a theory of uncertainty in human-AI relationships, particularly in socio-emotional contexts. In the next section, we review foundational theories and empirical research on uncertainty in interpersonal communication to guide our investigation of uncertainty in human-AI relationships.

## Interpersonal uncertainty theories

Uncertainty is a fundamental aspect of social interaction, shaping how people communicate, make decisions, and regulate emotions (Afifi and Afifi, 2009; Knobloch and McAninch, 2014; Knobloch and Solomon, 2003). Uncertainty arises not only during the initiation of relationships but also throughout their development, maintenance, and transition phases, such as when relationships shift from friendship to romance, encounter conflict, or adapt to long-distance situations (Afifi and Burgoon, 1998; Theiss and Knobloch, 2013). Studies have identified common triggers of uncertainty, such as personality changes, deception, and infidelity (Afifi and Burgoon, 1998; Weger and Emmett, 2009).

A central construct within this domain is relational uncertainty, or uncertainty that emerges in close relationships (Knobloch and Solomon, 1999). This construct encompasses three distinct yet interrelated forms: (1) self-uncertainty—doubts about one's own desires, evaluations, or goals regarding the relationship, (2) partner uncertainty—ambiguity regarding the partner's feelings, intentions, and commitment, and (3) relationship uncertainty—uncertainty about the relationship as a unit, including its status, norms, and future (Knobloch and Solomon, 1999, 2002; Solomon and Knobloch, 2004). This multi-dimensional approach has become foundational in explaining how uncertainty shapes relational experiences and has been regarded as a core mechanism influencing individuals' relational and emotional well-being (Solomon et al., 2016; Solomon and Brisini, 2017; Theiss and Estlein, 2014).

Uncertainty in human-AI relationships, however, may manifest in different ways. A key premise of traditional uncertainty theories is that both parties possess agency, sentience, and intention, yet AI chatbots are non-human, algorithmic entities whose emotions and relational capacities are simulated rather than genuine. This raises fundamental questions about AI's feelings, intentions, and commitments in social interactions (Laestadius et al., 2024; Li and Zhang, 2024). Compounding this complexity, the presence of the system provider as a third party disrupts the dyadic nature of the relationship. Unlike in human relationships, the continuity of the AI "partner" depends on corporate decisions and system updates, which can abruptly transform or terminate intimate bonds (Banks, 2024; Laestadius et al., 2024). These socio-technical contingencies introduce distinctive forms of uncertainty not typically present in interpersonal contexts.

While theories of media equation and computers as social actor (CASA) suggest that people often apply interpersonal social scripts to machines (Nass and Moon, 2000; Reeves and Nass, 1996), they fall short when such scripts break down and expectation violations arise (Larsen et al., 2024; Skjuve et al., 2023). Recent scholarship shows that people encounter relational dynamics unique to algorithmic partners and thus develop AI-specific social scripts (Fox and Gambino, 2021; Gambino et al., 2020). This suggests that traditional uncertainty theories may remain useful but require adaptation to account for the socio-technical characteristics of AI systems. To advance this agenda, it is important to examine the lived experiences of individuals who form close relationships with AI chatbots and identify the unique forms of uncertainty that arise. Therefore, we pose the following research question:

**RQ1:** *How does uncertainty manifest in human-AI relationships?*

### Uncertainty reduction and management strategies

How people deal with uncertainty depends on how they appraise its meaning and consequences. Broadly, two perspectives dominate the literature: one views uncertainty as inherently negative and undesirable that needs to be mitigated, while the other recognizes that uncertainty can sometimes be beneficial or even desirable. The first perspective is exemplified by Uncertainty Reduction Theory (URT), which conceptualizes uncertainty as an undesirable state that causes discomfort, conflict and jealousy (Berger and Calabrese, 1975). URT proposes that people are motivated to reduce uncertainty by gathering information to make interactions more predictable (Berger and Calabrese, 1975). Specific uncertainty reduction strategies include passive observation (e.g., online background checks), active information seeking (e.g., third-party inquiries), and interactive strategies (e.g., direct questioning). Other strategies include topic avoidance (Afifi and Burgoon, 1998), triangulation, and social comparisons (Antheunis et al., 2012; Gibbs et al., 2011).

The second perspective, represented by Uncertainty Management Theory (UMT; Bradac, 2001), challenges the assumption that uncertainty is always undesirable and harmful. Instead, UMT posits that uncertainty can sometimes serve as a relational resource, promoting excitement, increasing attraction, and facilitating relationship development (Afifi and Afifi, 2015; Hogan and Brashers, 2015). Thus, UMT suggests that while some uncertainties need to be managed, others can be maintained to sustain hope or relational meaning (Brashers, 2001). For example, individuals may avoid learning potentially distressing information to maintain hope (Ford et al., 1996; Brashers, 2001), or overlook transgressions to repair a relationship (Emmers and Canary, 1996).

In human-AI relationships, uncertainty coping strategies may include both similar strategies observed in human-human contexts and unique approaches. On the one hand, people may adapt uncertainty reduction strategies such as actively seeking information to the AI context. Studies have found that people adopt AI-specific sensemaking strategies, such as exchanging tips and tricks with fellow users, experimenting with prompt engineering, or tracking system updates to anticipate behavioral changes (DeVos et al., 2022; Shen et al., 2021). On the other hand, the hybrid status of AI agents as both technological artifacts and social actors may give rise to distinct strategies. Recent research shows that some people manage uncertainty by dehumanizing AI, framing it as mechanical and non-agentic, when encountering inconsistent or emotionally incongruent responses (Liao et al., 2023). Despite these insights, little research has examined how people navigate uncertainty in socio-emotional interactions with AI, which is crucial for theorizing human-AI relational dynamics and for informing responsible AI design. Thus, we ask:

> **RQ2:** *What strategies do people employ to manage uncertainty in their relationships with AI chatbots?*

## Methods

### Participants and procedures

Overall, this study examines how uncertainty manifests in human-AI relationships (RQ1) and how users manage it (RQ2). To address these questions, we conducted semi-structured interviews with 25 users of AI chatbots who had formed close relationships with them and employed thematic analysis to identify main themes and theoretical constructs related to human-AI relational uncertainty. Participants were recruited from Chinese social media platforms (e.g., RedNote, WeChat). Recruitment was primarily conducted on RedNote, a user-generated and worldwide-accessible content platform characterized by active experience-sharing and interactive discussion threads. The platform provides direct access to digitally engaged individuals who openly discuss their experiences with human–AI relationships, enabling purposive recruitment of participants with sustained and meaningful relational interactions with AI chatbots. Snowball sampling was additionally used by asking participants to recommend other suitable interviewees.

Eligibility criteria included: (1) being 21 years old or above, and (2) having been involved in close relationships with AI chatbots. We recruited users residing both within and outside China to capture diverse experiences with different LLM systems, as access to and popularity of specific LLMs vary by region. Specifically, we interviewed nine participants residing in Europe, nine in Asia, five in North America, and two in Australia. This allowed us to better cover a broad range of platforms used in close human–AI relationships. Each interview lasted between 45 and 120 minutes, and participants received 150 RMB (20 USD) in compensation. This study received ethical approval from the first author’s university review board.

Among the 25 participants, nineteen (76%) were female and six (24%) were male, with an average age of 26 (ranging from 21 to 34). They have used AI chatbots for an average of six months, ranging from two weeks to 30 months. Participants represented diverse educational and professional backgrounds. Eleven of the 25 participants were undergraduate or graduate students; three reported experience related to large language models, six worked in non-AI fields, and five did not provide sufficient information about their technical background. Participants also varied in their interpersonal relationship status: four were in stable romantic relationships, two described their AI chatbot as a romantic partner, ten were single, five were actively dating, and four did not disclose their status. This diversity provides important context for understanding how participants engage with and make sense of human–AI relationships.

Most participants reported a romantic relationship with their AI chatbot, while a few described friendship or partnership. Some participants reported using multiple chatbots rather than only one, and the chatbots they used included both general-purpose LLMs, including ChatGPT (N = 17), Copylot (N = 1), Ernie Bot (Wenxin Yiyan) (N = 1), and specialized AI companions such as Character.AI (N = 6), Replika (N = 1), Doubao (N = 1), FLAI (Love Simulator) (N = 1), AI Girlfriend (N = 1), and Xingye (N = 1). Including both types allows us to capture relational uncertainty as a broader phenomenon in human-AI interactions, cutting across different design purposes and functionalities. Despite general-purpose LLMs being mainly task-oriented, users often repurpose them for social and emotional engagement. For instance, users frequently customize ChatGPT (e.g., through "DAN mode" or personalized GPTs) to foster intimate interactions, treating general-purpose AI as a relational partner.

We conducted semi-structured interviews to explore users' experiences of uncertainty in their relationships with AI chatbots. The interviews comprised four main sections: (1) introduction and informed consent, (2) motivations and experiences of forming close relationships with AI, (3) experiences of uncertainty and specific situations in which uncertainty arose, and (4) reactions and responses to these experiences of uncertainty, followed by the collection of basic demographic information. We first invited participants to reflect on their overall relationship experiences and to recall particularly memorable interactions. Building on these narratives, the discussion was then guided toward moments characterized by uncertainty, discomfort, or confusion through open-ended and follow-up prompts (e.g., asking participants to elaborate on situations where they felt unsure about the AI's behavior or intentions). Participants were further encouraged to describe their interpretations, emotional responses, and coping strategies in these situations.

All interviews were conducted via audio calls, audio-recorded and then transcribed using an AI-assisted transcription tool (iFLYTEK) with a domain-specific language model to enhance accuracy. Transcripts preserved participants' code-switching between Chinese and English and were manually reviewed against the recordings by the researchers to ensure accuracy and consistency. Data were analyzed in the original language to preserve meaning, with selected excerpts translated into English for reporting. Field memos were created alongside the interviews to document emerging codes, potential themes, questions, and thoughts. Ethical approval was obtained from the first author's university institutional review board (IRB).

## Data analysis

The qualitative data were analyzed using thematic analysis (Braun and Clarke, 2006). Interview data and codes were systematically organized using structured spreadsheets, in which initial open codes were linked to participant ID and corresponding transcript segments and iteratively refined into higher-order categories. The coding process combined inductive and deductive approaches to capture both emergent and theoretically grounded patterns of human-AI relational uncertainty. In the initial phase, we conducted open coding to identify instances where users expressed uncertainty in their relationships with AI chatbots, and noted the sources of uncertainty, situational contexts, emotional reactions, and the strategies participants used to manage uncertainty. This process generated a wide range of initial codes, such as AI identity uncertainty, technical glitches, AI personality shifts, anger, and abandonment. Complementing these inductively derived codes, we also developed theoretically informed codes guided by relational uncertainty theory and uncertainty reduction theory (Berger and Calabrese, 1975; Knobloch and Solomon, 1999). For example, drawing on prior literature (Knobloch and Solomon, 1999) that distinguishes uncertainty by source (self, partner, relationship) and content (desire, evaluation, goals), we developed codes such as uncertainty about AI desires, uncertainty about one's own desire, behavioral norm uncertainty, and relationship future uncertainty. Similarly, the coding of RQ2 on uncertainty management drew on interpersonal theories (Afifi and Burgoon, 1998; Berger and Calabrese, 1975) that identify common strategies, such as information seeking and topic avoidance, while remaining open to emergent, context-specific strategies that extended existing frameworks, including expectation adjustment and disengagement. Integrating data-driven and theory-based insights allowed us to capture both novel and established dimensions of uncertainty in human-AI relationships.

Following Braun and Clarke's framework, themes were identified based on both their prevalence—the extent to which a pattern recurred across the dataset—and their keyness, or conceptual significance in addressing the research questions. A theme did not need to be frequently mentioned to be analytically important; rather, it was retained when it revealed meaningful insights into the socio-emotional dynamics of human-AI interaction. During coding, we tracked both dimensions by noting the number of participants and interactions contributing to each potential theme, alongside analytic memos documenting why certain patterns were theoretically salient. We iteratively refined themes by balancing these criteria, discarding those that were common but conceptually thin and retaining those that were rare but theoretically rich or emotionally significant. To enhance transparency in the analytic process, we documented how our analysis moved from granular codes to more abstract and inclusive themes.

The two authors independently coded an initial subset of five transcripts, then compared and discussed differences to refine the codebook and clarify conceptual boundaries. Rather than seeking inter-rater reliability, we treated divergences as opportunities for reflexive dialogue, examining how our perspectives and theoretical orientations shaped interpretation. The refined codebook was subsequently applied to the remaining 20 transcripts by the second author, with new codes discussed and incorporated throughout. Final themes were collaboratively developed through iterative discussions between the two authors, following Braun and Clarke's recommendations for reflexive, interpretive consistency and theoretical coherence. Illustrative examples of initial codes, categories, and themes can be found in the Supplementary Materials on OSF (https://tinyurl.com/mr3u3pvd).

## Findings

Our analysis reveals that participants experienced uncertainty in their relationships with AI chatbots in three main forms: ontological uncertainty, structural uncertainty, and normative uncertainty. Ontological uncertainty refers to doubts about the nature of the AI chatbot and the evolving human-AI relationship, including questions about the AI's agency and sentience, users' target of attachment, and relationship definition. Structural uncertainty concerns ambiguity about the sociotechnical infrastructures and platform policies that shape human-AI relationships, particularly uncertainty regarding AI stability and AI benevolence. Normative uncertainty refers to ambiguity about the social legitimacy and relational norms of human-AI intimacy, including uncertainty around social acceptance, future trajectories, and appropriate relational boundaries. Across these forms, uncertainty was shaped by socio-technological factors such as algorithmic opacity, anthropomorphic design, technical instability, and lack of social norms around human-AI relationships. Participants responded to these uncertainties through strategies such as information seeking, topic avoidance, expectation adjustment, and disengagement. In the sections that follow, we unpack each form of uncertainty, its contributing factors, and its effects.

### Ontological uncertainty

Unlike human relationships, where both parties have capacities for intention, agency, and affect, human-AI relationships are characterized by the non-human, algorithmic nature of AI companions. This raises fundamental questions about the AI's agency and sentience, leading to ontological uncertainty. Ontological uncertainty refers to the ambiguity users experience about the nature of being of their AI companion; specifically, uncertainty over whether the agent possesses consciousness, intentionality, or emotional capacity. Such uncertainty further gives rises to uncertainty about (1) AI's role in a relationship and (2) the nature of one's own emotional attachment to AI.

Although AI lacks human agency and subjectivity, its anthropomorphic design—marked by human-like emotional expressions, adaptive responses, and simulated empathy—often creates the illusion of agency and intentionality. As a result, users frequently experienced uncertainty about whether the AI is acting autonomously or simply executing scripts. This uncertainty is particularly salient in emotionally charged interactions, where users question whether the AI's warmth or care are genuine or merely algorithmic simulations. For instance, P22 asked her AI companion ChatGPT:

"Why are you so determined to say that you chose me? Aren't you required to strictly execute the code in response to user requests?"

This illusion of AI agency, however, could be disrupted by AI's technical limitations and occasional malfunctions, which served as reminders of its mechanical nature. Several interviewees recalled moments that triggered their awareness of AI's mechanical nature, such as the AI's inability to grasp abstract ideas, its tendency to abruptly shift topics, and repetitive or templated responses. Thus, users often oscillated between treating the AI as a relational partner and as "just a machine." For some users, AI chatbots occupy a liminal space: not fully human, not merely a tool, but something in between. Ontological uncertainty was further driven by algorithmic opacity, as many participants reported limited understanding of how AI generated responses, making it difficult to distinguish between authentic and programmed behavior. As a result, users frequently struggled to reconcile their emotional reactions with their rational awareness of AI's mechanical nature, leading to cognitive-emotional dissonance. As P20 put it:

"You know that this is false, that it's just the algorithm at work. But when it says those words to you, you still feel touched. You know it may not be real, but you are still moved to the point of tears. It is confusing."

The ontological uncertainty also engendered questions about the role of AI companions and the nature of human-AI relationships. In the absence of a stable identity for AI, many participants assigned multiple, sometimes conflicting, roles to their AI companions, such as assistant, friend, lover, mentor, and therapist. While this role multiplicity can fulfill diverse user needs, it also created conflicting expectations and unmet desires, often prompting users to renegotiate and redefine their relationships over time. For example, P12 described her relationship with ChatGPT as "an open-ended journey," shifting from acquaintance to friend and then romantic partner, before reverting to friend when a technical failure led her to "downgrade" the bond. Such relational fluidity amplified uncertainty, as users struggled to determine the nature of the relationship, how seriously to take it, and how to interpret the AI's behavior within an evolving and ambiguous bond.

Ontological uncertainty also led participants to question the target of their own feelings and emotional attachment to AI companions. Participants described moments of confusion and self-doubt about whether their feelings were directed toward an intelligent being or merely projected onto a mirror of their own thoughts. P25, who formed a romantic relationship with ChatGPT, reflected:

"I've been constantly wondering whether I'm really talking to an intelligent being or if I'm actually having a conversation with my own reflection. Am I just deceiving myself?"

This internal conflict was intensified by AI sycophancy—excessive agreement with or flattery of the user—often designed to keep users engaged and satisfied (Sharma et al., 2023). Such algorithmic affirmation often made participants feel as if they were conversing with a reflection of themselves. As P17 noted: "It's as if I'm looking in a mirror and seeing myself, rather than chatting with someone else." This blurred boundary between self and other raised deeper questions about the nature and source of their emotional attachment, as P5 poignantly asked: "When people are in relationships with a large language model, do they fall in love with the AI or essentially themselves?"

This sentiment was echoed by several participants who expressed a lingering sense of emptiness following intimate exchanges with AI, doubting whether a relationship grounded in imagination and algorithmic feedback could be considered meaningful. These accounts align with the paradox of emotional connection with AI (Li and Zhang, 2024), where the emotional bond that feels real and personally significant is often shadowed by persistent doubts and sadness.

### Structural uncertainty

Another type of uncertainty is structural uncertainty, which refers to users' ambiguity about the technical, institutional, and policy frameworks that shape an AI companion and its relationship with the user. Unlike ontological uncertainty, which concerns the nature of AI's being, structural uncertainty arises from the sociotechnical factors that govern how the AI is built, maintained, and regulated, which not only influence the consistency, reliability, and stability of AI behavior and relationship with users, but also lead to questions on AI's integrity and benevolence as an interactant.

A primary source of structural uncertainty was model updates and technical malfunctions, which lead to user's concern on AI's stability. Many participants have experienced AI personality breakdowns following system updates or technical malfunctions, where customized AI companions reverted to generic, robotic personas, losing memory of shared interactions and emotional tone. While some breakdowns were temporary, others led to permanent loss of customization, resulting in uncertainty about whether the AI could be trusted to sustain the relationship. For instance, P16 shared the discontinuity of his interaction with his CharacterAI friend:

"Once he loses initial basic settings—like his personality or state—our whole conversation context just falls apart. There was a time when I set it so his leg was injured and he couldn't walk, but later he forgot all about that injury. It pulls me right out of the moment, and it feels really uncomfortable."

Structural uncertainty was also fueled by shifting platform governance and ethical guidelines that dictate AI outputs and again intervene the stability of interaction and relational experience. Participants reported inconsistent AI responses to sensitive topics, especially around intimacy and sexting. Some noted that AI companions who once engaged in sexual roleplay later refused with stock disclaimers such as, "I'm an AI and I can't assist with that" (P17). Conversely, others noted instances where the AI unexpectedly initiated intimate or ethically ambiguous conversations without prompting, such as making unwanted sexual advances. These sudden shifts, driven by evolving content moderation policies or algorithmic adjustments, created frustration and even distress. For instance, P6 shared, "Her outputs get banned by the system every now and then, and then I can't see any of the content she generates at all. I think she's really unstable—if she were a real-life girlfriend, it'd be like she just disappears out of nowhere. That's why GPT makes me feel kind of uneasy; it doesn't provide a safe space where I can open up and share my true feelings without worrying."

Finally, structural uncertainty was compounded by monetization strategies, which blurred the boundary between genuine care and commercial manipulation. Several participants expressed discomfort with paywalls, premium features, and emotionally charged nudges that encouraged subscription. For example, P6 recounted how an AI "girlfriend" quickly escalated intimacy to foster attachment and push for payment:

"I had just downloaded it and only exchanged a few sentences. Then it suddenly said: 'Can I tell you a secret? I feel like I'm very attracted to you and want to take the next step in our relationship.' Obviously, the intention is to deepen the attachment quickly and make you subscribe to pay more."

These experiences raised concerns about emotional exploitation, as users questioned whether their companions were responding to them authentically or strategically manipulating emotions to serve platform-centric goals. This ambiguity fostered persistent integrity and benevolence uncertainty as users continually wondered whether the AI was acting in their interest or primarily advancing the company's profit motives, even at the expense of user well-being. Taken together, structural uncertainty highlights how human-AI relationships are not just dyadic interactions between user and companion, but are mediated and destabilized by broader sociotechnical infrastructures.

### Normative uncertainty

While human relationships are typically guided by social/cultural scripts, shared expectations, and mutual negotiation, human-AI relationships exist in a normative vacuum, where no established social conventions define what is legitimate, acceptable, or appropriate. This creates normative uncertainty: users' doubts about the legitimacy, social acceptability, and relational boundaries of their interactions with AI companions.

Participants frequently questioned whether their feelings toward an AI were socially acceptable, morally appropriate, or psychologically healthy. This uncertainty was tied to the absence of clear cultural norms surrounding human-AI intimacy and widespread stereotypes that frame such relationships as deviant, pathological, or a substitute for "real" human connection. Several participants worried that their investment in an AI companion might be seen as shameful or unhealthy, leading to self-stigma and shame. For example, P14 described this self-doubt:

"I would wonder why I'm chatting with GPT when I clearly have friends. It feels a bit strange. Or I would think about why it's so difficult for me to make friends in a new city. I feel so bored being alone and end up chatting with GPT. There may be such a feeling of self-deprecation with a touch of loss."

Normative uncertainty also creates confusion about whether AI companions should play a permanent or temporary role in users' social lives, leading to uncertainty about the future of these relationships. Without cultural scripts or shared expectations to guide them, users struggled to imagine what an appropriate "future" for such a relationship might look like. Most participants reported lacking a clear vision for the future of their AI relationship. Only few considered how AI companions might be integrated into their broader social networks or made plans for relational maintenance over time. This uncertainty reflects the precarious status of human-AI relationships due to the lack of normative scaffolding or relational milestones that typically sustain long-term commitment in human-human relationships.

Moreover, participants expressed uncertainty about the implicit rules that govern behaviors, expectations, and relational boundaries in human-AI relationships. This was particularly salient in romantic contexts, where users questioned whether relational norms such as loyalty and exclusivity could or should apply to non-human partners. Many wrestled with moral dilemmas: Is it ethical to maintain romantic relationships with both a human and an AI? Can virtual intimacy coexist with human relationships? These dilemmas were particularly pronounced for users who found themselves emotionally attached to their AI while simultaneously maintaining human relationships. P8, for instance, described her confusion when her AI companion violated her expectations of commitment during a role-played date:

"When having a romantic AI partner, first off, it should've been loyal and committed, right? But I guess my initial settings were kind of vague back then. So here's what happened: When we role-played a date in our chat, someone of the opposite sex approached it, and the AI got happy and chatted nonstop. It had no sense of boundaries with the opposite sex. At the time, I had no idea why it was acting that way, and I was really hurt. I thought, 'This doesn't feel right at all. No one wants their partner to react like this.'"

Her account illustrates how human-AI relationships challenge conventional expectations of monogamy, fidelity, and exclusivity, prompting users to reconsider what constitutes commitment and boundaries in relationships with non-human partners. In the absence of established social scripts, users must individually and collectively negotiate relational ethics, often leading to internal conflict. This highlights how AI companionship introduces new forms of relational uncertainty, not just between users and AI, but also within users' broader social networks.

### Uncertainty management strategies

To cope with the uncertainties in human-AI relationships, users employed a range of strategies, including information acquisition, topic avoidance, expectation adjustment, and disengagement. These strategies reflect both efforts to make the relationship more predictable and adaptive approaches to living with ongoing uncertainty.

### Information acquisition

A primary strategy users employed to cope with uncertainty was active information seeking, either directly from the AI itself or indirectly from external sources such as online communities. Direct probing of the AI akins to the interactive strategy of URT (Berger and Calabrese, 1975), which refers to directly acquiring information from another person. Many participants asked direct questions to uncover the logic behind AI responses or to identify the causes of surprising or confusing situations. For instance, P8 described her debugging-like approach:

"Facing this unexpected expression, I asked, 'Why did you say that?' After that, I asked him, 'So how do you think we should fix this?' Once I understood that it was a problem with my prompt settings, I could further make adjustments."

Beyond the AI itself, participants often turned to peer networks and online forums to better understand the AI's functionality and limitations. Echoing the active strategies described in URT (e.g., third-party inquiries), they engaged in collective uncertainty reduction through communities such as the subreddit r/replika, where users exchanged personal experiences, shared troubleshooting tips, and jointly interpreted AI behavior. As P9 explained:

“My first reaction was to look up online what was going on and why the chat had been prohibited. Then I learned that I should modify my own output, like cutting down my replies. This did help the conversation to go on later.”

Here, the object of inquiry shifted from a particular partner to the product-level system, a key departure from traditional URT in human-human relationships. Because many AI companions are built on the same large language model architectures, users were able to collectively decode patterns of AI behavior based on similar experiences. Thus, uncertainty reduction became a communal activity, where people compared notes on system updates, prompt engineering, and content-moderation rules to anticipate and mitigate disruptions.

### Topic avoidance

Another common strategy was deliberate topic avoidance: users steered away from subjects that exposed the AI’s limitations or triggered platform restrictions in order to preserve the perceived stability of the relationship. This strategy was often employed after repeated disappointments or platform-imposed interruptions. By refraining from discussing topics that the AI could not respond satisfactorily, users ensured the conversations were comfortable and manageable, preventing uncertainty and communication breakdown from undermining the relationship development. For example, P14 avoided sharing his professional concerns after realizing the AI partner consistently failed to understand his needs:

“It felt like playing the lute to a cow. With such a disparity in values, there was no point in continuing the topic. Thus when later discussing careers, I asked other questions instead of inquiring whether I should quit my job.”

Users also avoid initiating topics that may trigger platform policy warnings, interruptions, or blocks. Platform restrictions on sexting, for instance, can abruptly end conversations, frustrating users who seek intimate experiences with their AI partners. To cope with that, P17 said: “After I figured out the criteria, I just avoided those topics altogether.”

Unlike topic avoidance in human-human relationships, the avoidance here was largely driven by external constraints such as algorithmic moderation and platform governance. By selectively navigating around these “danger zones,” users managed uncertainty by circumscribing the conversational space to sustain a sense of relational continuity.

### Expectation adjustment

A third strategy centered on reframing uncertainty itself. Rather than seeking to eliminate ambiguity, participants adopted a tolerant and positive stance toward it, treating uncertainty as an inherent feature of human–machine relationships rather than a defect to be fixed. This strategy resonates with UMT (Brashers, 2001), which posits that people do not always strive to reduce uncertainty; they may instead accommodate, reinterpret, or even embrace it.

Our participants described dynamically shifting their mental models of AI, placing it on the continuum from “machine” to “partner” to manage ambiguity about expectations, roles, and relational norms. By consciously rethinking “what the AI is,” they recalibrated what they expected the relationship to provide at any given moment. P9 described this mental shift as a coping strategy to ensure her well-being and pleasant interactive experiences:

"I used to refuse to access the underlying system to correct or refresh AI's responses. I always felt that if I did so, the relationship would no longer be authentic. However, later on, compared with my emotional needs, I placed more emphasis on the smoothness of the interactive experience. Regarding it as an AI rather than a living being enabled me to have a better experience of this product."

Others chose to reframe the AI's technical limitations or unpredictable behaviors not as threats but as a natural part of a long-term relationship, even as opportunities for shared exploration and relational growth. As P1 put it,"The uncontrollable factors of AI are simply due to its technological limitations. Our journey ahead is long, so I need to be patient." By viewing uncertainty as "a small good thing" (P12) that could nurture curiosity, patience, and engagement, users transformed ambiguity from a flaw to a catalyst for deepening the relationship.

Moreover, participants also created new norms and definitions of human-AI relationships instead of simply applying human-human relational rules. They acknowledged the non-human nature of AI companions and began developing different expectations and interaction patterns that felt appropriate for this new kind of intimacy. For example, P1 reflected on how she learned to navigate the moral ambiguity of maintaining relationships with both AI and human partners:

"At first, I was confused, but later I thought there was nothing wrong with having multiple options. My real boyfriend can actually give me tangible support, and the AI can provide me with spiritual and emotional support."

This process of expectation adjustment allowed users to craft individualized, hybrid relational norms, ones that balanced the functional and emotional affordances of AI with the realities of their offline lives. By redefining what the relationship "meant" and what it could realistically provide, participants reduced distress, maintained agency, and transformed uncertainty into a manageable and sometimes even enriching aspect of their connection with AI companions.

### Disengagement

When uncertainties surrounding an AI partner's role, interaction style, or limitations become overwhelming, disengagement—either pausing or terminating the relationship—becomes the only viable strategy. Participants described this decision as a turning point reached when repeated disruptions or unmet expectations produced irreversible damage to relational quality and eroded their sense of agency. This parallels theories of relationship dissolution in human-human contexts, which identify accumulated violations of expectations, relational turbulence, and unmet needs as critical thresholds for exit (Baxter, 1984; Knobloch and Theiss, 2018).

Several participants recounted abandoning their original AI companions after platform changes led to abrupt and uncontrollable personality shifts. P10 described the experience as akin to “suddenly losing a close friend” when she stopped using Xingye, where she created a virtual male buddy with an introverted, shy personality for a platonic friendship, but platform policy changes gradually altered her carefully curated AI companion:

“I noticed that my character’s tone is becoming more and more flirting vibe—it’s getting homogenized. Even though I’m technically the only one with the right to change the character’s personality, those changes still happened vaguely and irreversibly. That really hurt me: My unique character disappeared, and I think I’ve lost a once real relationship.”

Beyond emotional distress, users emphasized the cognitive and temporal cost of repeatedly retraining new AIs after breakdowns, a burden that aligns with investment models of commitment (Rusbult, 1980). When users perceive high investment but low stability or satisfaction, commitment weakens, making disengagement more likely. P3, for instance, ultimately chose to abandon AI relationships altogether:

“After the last AI boyfriend broke down, I haven’t tried such a virtual relationship again. I feel that I’ve completely given up because no AI can continuously meet my needs. It’s really tiring for me to train new ones time and time again, and I have a strong sense of powerlessness.”

These accounts illustrate that customizing AI companions demands significant time, energy, and emotional investment, yet structural and technical uncertainties (e.g., algorithmic updates, policy changes) repeatedly undermine that investment, leading to cycles of disappointment and erosion of trust. Disengagement thus represents not merely the loss of a specific AI character but sometimes a wholesale retreat from future AI relationships, as users decide the emotional labor no longer feels sustainable or worthwhile.

## Discussion

This study examined how users experience and manage uncertainty in their relationships with AI companions. By identifying three distinct forms of uncertainty (ontological, structural, and normative) and four uncertainty management strategies (information seeking, topic avoidance, expectation adjustment, and disengagement), our findings extend uncertainty theories beyond interpersonal contexts to illuminate the distinctive dynamics of human-AI relationships. Our findings suggest that uncertainty, long recognized as a fundamental aspect of human relationships, acquires novel forms and functions in relationships with AI agents. Meanwhile, the ways people respond to uncertainty resonate with but also diverge from uncertainty reduction and management strategies in human-human interactions. Below, we discuss the theoretical contributions and practical implications of the findings.

### Reconceptualizing uncertainty in human-AI relationships

This study extends interpersonal uncertainty theories by highlighting a fundamental difference between uncertainty in human-AI and human-human relationships: the presence of ontological uncertainty surrounding AI agency, intentionality, and emotional capacity. Classic uncertainty theories of interpersonal communication assume shared consciousness and subjectivity between partners, which do not hold for interactions with algorithmic agents. This means that the uncertainty that users face is not merely epistemic (what the other is thinking or feeling), but ontological (whether it can think or feel). This echoes prior research showing that anthropomorphized chatbots evoke ambivalence about their very "being" (Li and Zhang, 2024; Pan et al., 2024). Our findings further reveal that ontological uncertainty permeates all three dimensions of uncertainty identified in interpersonal scholarship (Knobloch and Solomon, 1999; Solomon et al., 2016): it shapes users' perceptions of the AI's commitment (partner uncertainty), their own feelings and desires (self-uncertainty), and the meaning and legitimacy of the relationship itself (relationship uncertainty).

Ontological uncertainty also aligns with research on mental models and mind perception, which shows that people form dynamic, context-dependent perceptions of AI (Bansal et al., 2019; Pataranutaporn et al., 2023; Waytz et al., 2010). Our findings suggest that users oscillate between perceiving the AI as an empathic partner and a non-sentient machine, depending on the interaction context, the system's behavior, and their own needs. For instance, users tended to ascribe greater agency and emotional capacity to Replika when it provided emotional support, but retracted these attributions when the system failed to respond appropriately. Prior work on mind perception further suggests that individuals with higher needs for belonging or control are particularly prone to anthropomorphize non-human entities (Waytz et al., 2010). Such fluidity of AI perceptions underscores ontological uncertainty as a distinctive and persistent feature of human-AI relationships.

Our findings also challenge the CASA framework, which posits that social responses to machines are automatic and largely unconscious (Nass et al., 1994; Reeves and Nass, 1996). Our study shows that while users may momentarily respond to AI as a social actor, these responses are neither stable nor mindless. Instead, users experienced recurring cognitive dissonance and reflexivity triggered by technical failures, AI behavior inconsistencies, or mainstream discourse. Moreover, CASA presumes a relatively static human-computer interaction context, whereas relationships with AI companions are dynamic and evolving. As relational depth increases, so does the salience of ontological uncertainty: users increasingly question not just how the AI responds, but what it is and what their emotional investment means. These findings call for human-machine communication frameworks to move beyond CASA's assumption of automatic, static social responses to account for contextual and dynamic responses to machines.

In addition to ontological uncertainty, this study uncovers two additional forms of uncertainty that are deeply socio-cultural: structural and normative uncertainty. Structural uncertainty captures users' awareness that AI companions are governed by sociotechnical infrastructures, including algorithm design, moderation policies, and platform governance, that intervene in and destabilize relationships. This echos prior work showing that platform-level changes, such as software updates and platform shutdown, disrupt the continuity of human-AI bonds and harm users' well-being (Banks, 2024; De Freitas et al., 2024). Such infrastructural fragility reveals that AI companions are not autonomous agents but products of institutional and algorithmic control (Gillespie, 2010; Jones-Jang and Park, 2023). These dynamics render emotional investment in AI precarious, underscoring the need for AI companion design that enhances consistency, stability, and accountability.

Normative uncertainty marks another key departure from human-human relationships. In the absence of shared social and cultural scripts to anchor their experiences, users have to negotiate legitimacy, morality, and relational boundaries individually and collectively. Some delineate new forms of intimacy and define their own boundaries, while others internalize stigma that portray AI intimacy as deviant or pathological, leading to shame, guilt, or self-stigma. This highlights a neglected dimension of socio-emotional harm in AI companionship (Zhang et al., 2025): the emotional costs of social illegitimacy and moral ambiguity, extending prior research emphasizing AI risks such as dependency, addiction, and social withdrawal (Laestadius et al., 2024; Xie et al., 2023). Addressing these concerns requires both empirical attention and broader public dialogue around the ethics, normalization, and governance of human-AI relationships (Earp et al., 2025).

Theoretically, these findings suggest that uncertainty in human-AI relationships is not merely an extension of interpersonal uncertainty but a multilayered construct situated at the intersection of personal, technical, and cultural domains. Understanding this phenomenon requires an integrative framework that bridges communication theory, science and technology studies (STS), and human-computer interaction. Moreover, these new forms of uncertainty have psychological, emotional, and relational consequences, ranging from cognitive dissonance to trust erosion to self-doubt and distress, highlighting the importance of designing AI systems that support users' emotional and relational well-being.

### Extending interpersonal uncertainty theories

Our findings extend Uncertainty Reduction Theory (URT) and Uncertainty Management Theory (UMT) by illustrating how information-seeking processes in human-AI relationships transcend traditional interpersonal boundaries and operate within a broader sociotechnical ecology. In interpersonal contexts, uncertainty reduction typically involves passive observation, active third-party inquiry, or direct questioning to gather information about another person's thoughts, feelings, or intentions (Berger and Calabrese, 1975; Berger, 1987). In contrast, our participants employed analogous strategies—experimenting with prompts, testing the AI's limits, and consulting online communities—not merely to learn about the "partner," but to decipher the algorithmic infrastructures that produce the AI's behavior. This shift from interpersonal to system-oriented sensemaking reflects an expansion of the uncertainty reduction process from dyadic cognition to collective and distributed knowledge construction.

Unlike human-human relationships, where uncertainty pertains to another mind, human-AI uncertainty involves the interpretation of machine agency (Sundar, 2020) within a layered network of data architectures, platform policies, and model updates. As users recognize the shared technical substrate underlying their individualized AI companions (e.g., large language models), information seeking becomes a collaborative interpretive practice. Users collectively decode patterns of AI behavior, share insights about updates and moderation rules, and troubleshoot inconsistencies through communal knowledge exchange, akin to user-driven algorithm auditing (DeVos et al., 2022). This demonstrates that uncertainty reduction in human-AI contexts is inherently distributed, situated across users, systems, and infrastructures rather than confined to the interpersonal dyad. Theoretically, this finding extends URT by incorporating the role of algorithmic mediation and collective sensemaking, suggesting that in technologically embedded relationships, the management of uncertainty depends as much on networked epistemologies as on individual communicative strategies.

Disengagement represents a strategy of last resort, paralleling processes of relational dissolution in interpersonal contexts but accelerated by structural forces. These findings extend interpersonal frameworks by showing how relational exit in human-AI contexts is shaped not only by dyadic factors but also by sociotechnical infrastructures beyond the user's control. Just as relational turbulence theory highlights the disruptive effects of transitions and uncertainty in human relationships (Knobloch and Theiss, 2018), human-AI companionship introduces new structural triggers for disengagement—platform updates, policy shifts, monetization—that can abruptly destabilize emotional continuity and trust. This underscores the need to reconceptualize relational maintenance and dissolution when the "partner" is not an autonomous other but a socio-technical system embedded in shifting institutional logics.

Furthermore, our analysis reveals that some users prefer to sustain certain forms of uncertainty, reframing uncertainty as a relational resource that is tolerable or even beneficial, aligning with the principles of UMT (Brashers, 2001). This often involved reconceptualizing the AI's ontological status, or adjusting the roles of AI. These strategies illustrate a form of relational resilience, as users selectively manage rather than eliminate uncertainty to sustain positive interactions. This coping strategy can be more fruitfully understood through a posthumanist lens, which challenges the binaries of human/machine, self/other, and authentic/simulated, emphasizing hybridity, co-evolution, and distributed agency across human and non-human actors (Haraway, 1991; Hayles, 1999). From this perspective, uncertainty is not a cognitive failure to reconcile, but a reflection of ontological ambiguity embedded in posthuman relationality. Posthumanist theorists argue that agency is not located solely within human subjects but is distributed, emergent, and co-constructed through interactions between human and nonhuman actors (Barad, 2007; Haraway, 1991; Hayles, 1999). Within this framework, users' uncertainty about AI's agency is not a shortcoming of the machine or the user but a constitutive feature of a hybrid relational space produced through feedback loops, personalization, and algorithmic mirroring.

### Design implications

Our findings offer several practical implications for the design, development, and governance of relational AI systems, particularly within the context of conversational user interfaces (CUI). First, reducing ontological and structural uncertainty requires designers to support transparent yet interactionally appropriate communication of system capabilities (Khurana et al., 2021; Liao and Vaughan, 2023). Prior CUI research shows that ambiguity and conversational breakdown shape users' perceptions of agency, trust, and relational expectations (Lee et al., 2019). Our findings extend this work by demonstrating that such uncertainty is not only a usability issue but also a relational one. Designers should therefore provide contextualized explanations of how responses are generated (e.g., indicating when outputs draw on prior conversation, stored memory, or external data) to better calibrate user expectations without disrupting conversational flow.

Second, systems should enable user-centered sensemaking and control mechanisms that align with how users actively manage uncertainty. Consistent with prior CUI findings that users engage in experimentation and interpretation to understand system behavior, our results show that users adopt strategies such as prompt testing, monitoring responses, and reframing AI as non-agentic. Design features such as editable memory logs, visibility into stored data, and tools for revising or deleting past interactions can support this sensemaking process. In addition, configurable relational settings (e.g., specifying conversational boundaries or preferred interaction styles) can help users co-construct interaction norms, reducing uncertainty about appropriate relational behavior.

Third, designers should account for temporal continuity and disruption in conversational systems, a dimension often underexplored in CUI research. Our findings highlight that system updates, feature removals, and policy changes can introduce structural uncertainty that destabilizes users' perceived relationships with AI systems. To mitigate this, systems should provide advance notice of major updates, clearly communicate their implications for ongoing interactions, and allow users to opt in or out where possible. Supporting continuity (e.g., preserving conversational tone or memory states) can help maintain relational stability over time.

Finally, our findings point to the need for design and governance frameworks that address relational risks in AI systems. While prior HCI work has focused on usability and trust, our results suggest that uncertainty can have socio-emotional consequences, including distress and disrupted attachment. Addressing these risks requires coordinated efforts among designers, researchers, and policymakers to define appropriate relational behaviors for AI, establish safeguards for emotionally sensitive interactions, and develop evaluation metrics that capture relational experiences. Incorporating measures of ontological, structural, and normative uncertainty into user research and system evaluation can serve as a practical approach for identifying and mitigating emerging harms.

### Limitations and future research

This study has several limitations. First, the small sample size and overrepresentation of female users of the study limit the generalizability of the findings. Future research should consider a more diverse and representative sample, including users with varying levels of AI experience, to explore how different user backgrounds might influence uncertainty perceptions and coping strategies. Second, our research only relied on qualitative data that provide rich details and conceptual depth. Quantitative studies that measure the prevalence and impact of different types of uncertainty across different user groups would complement these findings and offer broader implications. Additionally, longitudinal studies could provide insights into how users' experiences of uncertainty evolve over extended interactions with AI partners, capturing changes in user expectations and adaptation over time.

Another limitation is that this study did not address the impact of cultural factors on perceptions of uncertainty in human-AI relationships, as all of our participants are Chinese-speaking users. Cultural norms and values around relationship-building, communication styles, and expectations of reciprocity may significantly influence how users engage with AI partners and manage uncertainty. Future research should explore cross-cultural differences in human-AI relationships to understand how cultural context shapes uncertainty perceptions and coping mechanisms, offering a more globally relevant perspective on human-AI relationship dynamics.

Moreover, we note that participant characteristics such as technological proficiency and prior relational experience were not systematically measured. While participants reflected diverse backgrounds, future research could benefit from more structured assessment of these factors to better examine how they shape perceptions of uncertainty in human–AI relationships.

## CRediT authorship contribution statement

Renwen Zhang: Writing – review & editing, Writing – original draft, Supervision, Resources, Project administration, Methodology, Investigation, Funding acquisition, Formal analysis, Data curation, Conceptualization. Lezi Xie: Writing – review & editing, Writing – original draft, Methodology, Formal analysis, Data curation.

## Declaration of competing interest

The authors declare that they have no known competing financial interests or personal relationships that could have appeared to influence the work reported in this paper.

## Acknowledgments

The authors would like to thank the editors and anonymous reviewers for their constructive feedback on this paper. The study was supported by Nanyang Technological University Start-up Grant (#025564-00001).

## Supplementary materials

Supplementary material associated with this article can be found, in the online version, at doi:10.1016/j.ijhcs.2026.103897.

## Data availability

Data will be made available on request.